\documentclass[fleqn,usenatbib]{mnras}

\usepackage{newtxtext,newtxmath}
\usepackage[T1]{fontenc}
\usepackage{ae,aecompl}
\usepackage{color}
\usepackage{graphicx}	
\usepackage{amsmath}	
\usepackage{amssymb}	
\usepackage{threeparttable}

\title[GRB\,140713A]{Detailed multi-wavelength modelling of the dark GRB 140713A and its host galaxy}

\author[A. B. Higgins et al.]{A. B. Higgins,$^{1}$\thanks{E-mail: abh13@le.ac.uk}
A. J. van der Horst,$^{2,3}$
R. L. C. Starling,$^{1}$
G. Anderson,$^{4,5}$ \newauthor
D. Perley,$^{6}$
H. van Eerten,$^{7}$
K. Wiersema,$^{8}$
P. Jakobsson,$^{9}$
C. Kouveliotou,$^{2}$ \newauthor
G. P. Lamb$^{1}$ and
N. R. Tanvir$^{1}$
\\
$^{1}$Department of Physics and Astronomy, University of Leicester, University Road, Leicester, LE1 7RH, UK\\
$^{2}$Department of Physics, The George Washington University, 725 21st Street NW, Washington, DC 20052, USA\\
$^{3}$Astronomy, Physics, and Statistics Institute of Sciences (APSIS), 725 21st Street NW, Washington, DC 20052, USA\\
$^{4}$International Centre for Radio Astronomy Research, Curtin University, GPO Box U1987, Perth, WA 6845, Australia\\
$^{5}$Department of Physics and Astrophysics, University of Oxford, Denys Wilkinson Building, Oxford, OX1 3RH, UK\\
$^{6}$Astrophysics Research Institute, LJMU, IC2, Liverpool Science Park, 146 Brownlow Hill, Liverpool, L3 5RF, UK\\
$^{7}$Department of Physics, University of Bath, Claverton Down, Bath, BA2 7AY, UK\\
$^{8}$Department of Physics, University of Warwick, Coventry, CV4 7AL, UK\\
$^{9}$Centre for Astrophysics and Cosmology, Science Institute, University of Iceland, Dunhagi 5, 107 Reykjavik, Iceland}

\date{Accepted XXX. Received YYY; in original form ZZZ}

\pubyear{}

\begin{document}
\label{firstpage}
\pagerange{\pageref{firstpage}--\pageref{lastpage}}
\maketitle

\begin{abstract}
We investigate the afterglow of GRB\,140713A, a gamma-ray burst (GRB) that was detected and relatively well-sampled at X-ray and radio wavelengths, but was not present at optical and near-infrared wavelengths, despite searches to deep limits. 
We present the emission spectrum of the likely host galaxy at $z = 0.935$ ruling out a high-redshift explanation for the absence of the optical flux detection. 
Modelling the GRB multi-wavelength afterglow using the radiative transfer hydrodynamics code \textsc{boxfit} provides constraints on physical parameters of the GRB jet and its environment, for instance a relatively wide jet opening angle and an electron energy distribution slope $p$ below 2. Most importantly, the model predicts an optical flux about two orders of magnitude above the observed limits. We calculated that the required host extinction to explain the observed limits in the $r$, $i$ and $z$ bands was $A^{\rm host}_{V} > 3.2$\,mag, equivalent to $E(B-V)^{\rm host} > 1.0$\,mag. From the X-ray absorption we derive that the GRB host extinction is $A^{\rm host}_{\rm V} = 11.6^{+7.5}_{-5.3}$\,mag, equivalent to $E(B-V)^{\rm host} = 3.7^{+2.4}_{-1.7}$\,mag, which is consistent with the extinction required from our \textsc{boxfit} derived fluxes. We conclude that the origin of the optical darkness is a high level of extinction in the line of sight to the GRB, most likely within the GRB host galaxy.
\end{abstract}


\section{Introduction}
Gamma-ray bursts (GRBs) are short-lived, explosive transients that can be detected at multiple wavelengths. The prompt gamma-ray emission is likely caused by internal shocks of material accelerated to relativistic speeds with a range of Lorentz factors \citep{Rees1994}. The afterglow, produced when the relativistic ejecta shocks into the surrounding medium \citep{Piran1999,Zhang2004,Kumar2015,Schady2017,vanEerten2018}, lasts longer and produces broadband emission ranging from X-ray to radio wavelengths. Observing the multi-wavelength emission of the afterglow can be used to study the interaction of a GRB with its environment.

Some GRBs detected at X-ray and radio wavelengths have lower than expected fluxes or deep limits in the optical bands. These are referred to as `dark' bursts, the earliest documented being GRB\,970828 \citep{Groot1998}. This suppression of optical flux can have a variety of causes. At high redshifts, the most likely cause for the optical darkness is Lyman-$\alpha$ absorption occurring at ${\rm\lambda_{obs}} < 1216(1+z)$\,\text{\AA} \citep{Tanvir2009,Cucchiara2011}. If the GRBs reside at lower redshifts, the observed optical darkness may result from a number of possibilities. Extinction due to line of sight dust contributions from the host, our Galaxy and the interstellar medium (ISM) can highly obscure the rest frame optical flux of GRBs. A previous investigation by \citet{Perley2009} observing 29 host galaxies of dark GRBs concluded that a significant fraction of hosts (six out of 22 with estimated dust extinction) had a moderately high level of extinction ($A_{\rm V}^{\rm host} > 0.8$). Furthermore, some dark bursts appear to reside in hosts with very high extinction (e.g., GRB\,111215A where $A_{\rm V}^{\rm host} > 7.5$; \citealt{Zauderer2013,VanDerHorst2015}). As GRBs trace cosmic star formation through their high energy emission \citep{Perley2016}, and a significant fraction of star formation is dust-obscured, dark GRBs may provide a way to investigate dust-obscured star formation \citep{Blain2000,Ramirez-Ruiz2002}. It is also possible that a GRB has either a low luminosity or low frequency synchrotron cooling break, and the subsequent afterglow would not have produced an optical flux that was detectable due to current instrument sensitivity or optical follow-up that simply was not deep enough. Coupled with a moderate extinction, many dark GRBs may not have been detectable in the optical at all.

The launch of NASA's Neil Gehrels Swift Observatory (\textit{Swift}; \citealt{Gehrels2004}) satellite in 2004 allowed rapid, follow-up observations of GRB afterglows. This has given us unprecedented multi-wavelength coverage of GRBs that was not possible previously. The percentage of GRBs found by the \textit{Swift} Burst Alert Telescope (BAT; \citealt{Barthelmy2005}) and additionally detected by the X-ray telescope (XRT; \citealt{Burrows2005}) stands at $> 90\%$ (\citealt{Burrows2008,Evans2009}; Swift GRB table\footnote{https://swift.gsfc.nasa.gov/archive/grb$\_$table/ \label{fn:grbtab}}). The \textit{Swift} Ultraviolet/Optical Telescope (UVOT; \citealt{Roming2005}) has detected an optical afterglow candidate in $\sim 30\%$ of \textit{Swift}/BAT detected GRBs (\citealt{Roming2009}, Swift GRB table$^{\ref{fn:grbtab}}$).

A number of investigations complementing \textit{Swift} XRT detections with optical follow-up have been undertaken (e.g., \citealt{Fynbo2009,Greiner2011,Melandri2012}). These samples differ in selection criteria but estimates have shown that dark bursts may account for $25-40\%$ of the \textit{Swift} GRB population.

Several methods have been proposed to classify these so called dark bursts, comparing the X-ray afterglow properties to the optical/nIR upper limits. \citet{Rol2005} estimated a minimum optical flux by extrapolating from the X-ray flux using both temporal and spectral information, assuming a synchrotron spectrum. \citet{Jakobsson2004} and \citet{VanDerHorst2009} characterise `thresholds' to classify a dark GRB using the optical-to-X-ray spectral index. These comparisons can only be used provided the observations used are made several hours after the GRB onset. \citet{Jakobsson2004} and \citet{VanDerHorst2009} highlight that classifications using spectral slopes alone may not fully determine whether a burst is truly dark, suggesting that these thresholds should only be used as quick diagnostic tools. They suggest that dark bursts should be modelled individually to fully characterise their nature. If multi-wavelength data are available (i.e. radio and X-ray), broadband modelling can be used to estimate the expected optical fluxes to determine the host galaxy optical extinction (discussed in \citealt{VanDerHorst2015}). Some dark GRBs with well sampled data at both the X-ray and radio wavelengths have been studied in detail; GRB\,020819 \citep{Jakobsson2005}, GRB\,051022 \citep{CastroTirado2007,Rol2007}, GRB\,110709B \citep{Zauderer2013} and GRB\,111215A \citep{Zauderer2013,VanDerHorst2015}. However, this sample is still small and highlights the importance to analyse new dark bursts to investigate the properties of the burst, the origin of the optical darkness and the use of dark GRBs as probes of dust obscured star formation. 

We investigate GRB\,140713A, a burst discovered by \textit{Swift} \citep{Mangano2014} and \textit{Fermi}/GBM \citep{Zhang2014}. GRB\,140713A was a long duration burst with a T$_{90} \sim 5$\,s ($15-350$\,keV) and a fluence $F_{\gamma} = 3.7(\pm0.3)\times10^{-7}$\,erg\,cm$^{-2}$ ($15-150$\,keV; \citealt{Stamatikos2014}). 
An X-ray counterpart was detected by the \textit{Swift}/XRT with initial localization uncertainty of 2\,arcsec (90\% containment; \citealt{Beardmore2014}) though this was later improved to 1.4\,arcsec\footnote{http://www.swift.ac.uk/xrt$\_$positions/} (90\% containment). A radio counterpart was also detected at 15.7\,GHz with the Arcminute Microkelvin Imager (AMI) Large Array \citep{Anderson2014a} coincident with the \textit{Swift}/XRT position. A potential host galaxy was found with the 10.4\,m Gran Telescopio Canarias (GTC; \citealt{CastroTirado2014}). We model the multi-wavelength afterglow data using numerical modelling based on hydrodynamical jet simulations - the first time this has been attempted on an optically dark GRB. The modelling can estimate the optical flux we would expect from the GRB and can be used to investigate the origin of the optical darkness.

\section{Observations and Data Analysis} \label{sec:obsanal}
\subsection{Radio observations} \label{sec:obsradio}

\begin{table*}
\caption{AMI and WSRT observations of GRB\,140713A where $\Delta T$ is the midpoint of each observation in days after the GRB trigger time. Non-detections are given as $3\sigma$ upper-limits. The AMI data are identical to those quoted in \citet{Anderson2018}.}
	\centering
	\begin{tabular}{|l|c|c|c|c|c|}
	\hline
	Epoch & $\Delta T$ & Integration & Observatory & Frequency & Flux \\ 
    & (days) & time (hours) & & (GHz) & ($\mu$Jy) \\ \hline
    Jul 13.784 - 13.867 & 0.05 & 2.0 & AMI & 15.7 & $<270$ \\
    Jul 14.791 - 14.958 & 1.09 & 4.0 & AMI & 15.7 & $<180$ \\
    Jul 16.884 - 17.050 & 3.18 & 4.0 & AMI & 15.7 & $600(\pm90)$ \\
    Jul 17.858 - 18.024 & 4.16 & 4.0 & AMI & 15.7 & $<270$ \\
    Jul 18.793 - 18.959 & 5.09 & 4.0 & AMI & 15.7 & $780(\pm90)$ \\
    Jul 19.687 - 20.185 & 6.15 & 12.0 & WSRT & 4.8 & $<96$ \\
    Jul 19.936 - 20.102 & 6.24 & 4.0 & AMI & 15.7 & $840(\pm70)$ \\
    Jul 20.943 - 21.109 & 7.24 & 4.0 & AMI & 15.7 & $820(\pm90)$ \\
    Jul 22.921 - 23.087 & 9.22 & 4.0 & AMI & 15.7 & $1370(\pm80)$ \\
    Jul 24.673 - 25.172 & 11.16 & 12.0 & WSRT & 4.8 & $189(\pm34$) \\
    Jul 24.860 - 25.027 & 11.16 & 4.0 & AMI & 15.7 & $1310(\pm100$) \\
    Jul 26.894 - 27.061 & 13.18 & 4.0 & AMI & 15.7 & $1650(\pm100)$ \\
    Jul 28.784 - 28.950 & 15.08 & 4.0 & AMI & 15.7 & $870(\pm70)$ \\
    Jul 30.657 - 31.155 & 17.11 & 12.0 & WSRT & 4.8 & $205(\pm28)$ \\
    Jul 30.807 - 30.973 & 17.11 & 4.0 & AMI & 15.7 & $690(\pm70)$ \\
    Aug 1.859 - 2.025 & 19.16 & 4.0 & AMI & 15.7 & $890(\pm70)$ \\
    Aug 3.860 - 4.026 & 21.16 & 4.0 & AMI & 15.7 & $1050(\pm70)$ \\
    Aug 5.815 - 5.981 & 23.12 & 4.0 & AMI & 15.7 & $700(\pm70)$ \\
    Aug 6.838 - 7.136 & 24.11 & 12.0 & WSRT & 4.8 & $137(\pm31)$ \\
    Aug 6.868 - 7.034 & 24.17 & 4.0 & AMI & 15.7 & $790(\pm60)$ \\
    Aug 7.635 - 8.133 & 25.10 & 12.0 & GMRT & 1.4 & $<225$ \\
    Aug 12.792 - 12.917 & 30.07 & 3.0 & AMI & 15.7 & $710(\pm70)$ \\
    Aug 14.871 - 14.995 & 32.15 & 3.0 & AMI & 15.7 & $530(\pm70)$ \\
    Aug 16.870 - 18.947 & 34.13 & 2.0 & AMI & 15.7 & $400(\pm60)$ \\
    Aug 18.605 - 19.103 & 36.07 & 12.0 & WSRT & 4.8 & $189(\pm32)$ \\
    Aug 18.781 - 18.947 & 36.08 & 4.0 & AMI & 15.7 & $490(\pm70)$ \\
    Aug 20.786 - 20.869 & 38.05 & 2.0 & AMI & 15.7 & $<180$ \\
    Aug 23.726 - 28.014 & 41.03 & 4.0 & AMI & 15.7 & $350(\pm50)$ \\
    Aug 27.848 - 28.014 & 45.15 & 4.0 & AMI & 15.7 & $290(\pm40)$ \\
    Aug 29.823 - 29.989 & 47.12 & 4.0 & AMI & 15.7 & $270(\pm50)$ \\
    Aug 31.757 - 31.832 & 49.01 & 1.8 & AMI & 15.7 & $<210$ \\
    Sep 1.795 - 1.962 & 50.09 & 4.0 & AMI & 15.7 & $320(\pm80)$ \\
    Sep 2.596 - 3.062 & 51.05 & 12.0 & WSRT & 4.8 & $182(\pm36)$ \\
    Sep 2.683 - 2.928 & 51.02 & 5.9 & AMI & 15.7 & $180(\pm40)$ \\
    Sep 5.754 - 5.919 & 54.05 & 4.0 & AMI & 15.7 & $<120$ \\
    Sep 7.778 - 7.942 & 56.08 & 3.9 & AMI & 15.7 & $<210$ \\
    Sep 10.798 - 10.965 & 59.10 & 4.0 & AMI & 15.7 & $210(\pm50)$ \\
    Sep 14.716 - 14.882 & 63.02 & 4.0 & AMI & 15.7 & $<150$ \\
    Sep 17.543 - 18.021 & 66.00 & 12.0 & WSRT & 4.8 & $192(\pm38)$ \\
    Sep 17.658 - 17.899 & 66.00 & 5.8 & AMI & 15.7 & $<90$ \\
    Sep 23.766 - 23.932 & 72.07 & 4.0 & AMI & 15.7 & $<120$ \\
    Oct 2.482 - 2.980 & 80.95 & 12.0 & WSRT & 4.8 & $127(\pm32)$ \\
    Oct 2.590 - 2.833 & 80.93 & 5.8 & AMI & 15.7 & $<150$  \\ \hline
	\end{tabular} 
\label{tab:data}
\end{table*}

We observed GRB\,140713A from 2014 July 13 to October 2 with the Large Array of the AMI interferometer \citep{Zwart2008} at a central frequency of 15.7\,GHz (between $13.9-17.5$\,GHz), and with WSRT at 1.4 and 4.8\,GHz. The AMI observations were taken as part of the AMI Large Array Rapid-Response Mode (ALARRM) program, which is designed to probe the early-time radio properties of transient events by automatically responding to transient alert notices \citep{Staley2013,Staley2016,Anderson2018}. On responding to the \textit{Swift}-BAT trigger of GRB 140713A, AMI was observing the event within 6\,min for 2\,hrs, obtaining a $3\sigma$ flux upper-limit of $0.27$\,mJy. Follow-up observations were manually scheduled and obtained every few days for over 2 months, with the first confirmed detection occurring 3.19\,d post-burst \citep{Anderson2018}. AMI data were reduced using the AMI\textsc{survey} software package \citep{Staley2015a}, which utilises the AMI specific data reduction software \textsc{ami-reduce} \citep{Dickinson2004} and \textsc{chimenea}, which is built upon the Common Astronomy Software Application (\textsc{CASA}; \citealt{Jaeger2008}) package and specifically designed to clean and image multi-epoch transient data \citep{Staley2015b,Staley2015c}. All flux densities were measured using the Low Frequency Array Transient Pipeline (\textsc{trap}; \citealt{Swinbank2015}) and the quoted flux errors were calculated using the quadratic sum of the error output by \textsc{trap} and the 5\% flux calibration error of AMI \citep{Perrott2013}. For further details on the reduction and analysis we performed on the AMI observations, see \citet{Anderson2018}.

\begin{figure*}
\centering
\includegraphics[width=\textwidth]{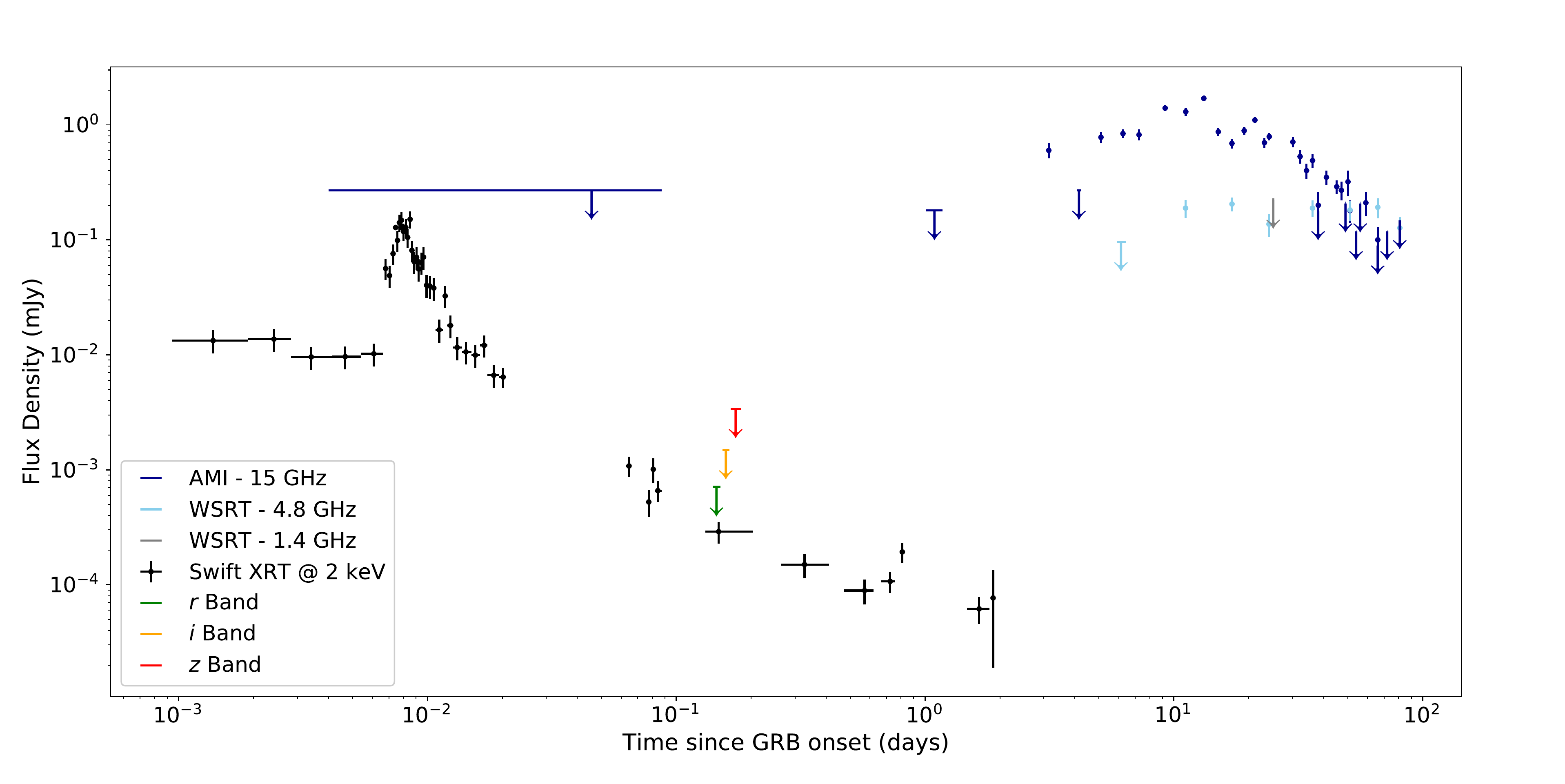}
\caption{Time evolution of GRB\,140713A at radio and X-ray wavelengths. The flux density detection errors quoted are $1\sigma$ and the non-detections are given as $3\sigma$ upper limits. The $r$, $i$ and $z$ optical upper limits are also plotted. The X-ray flux density is derived from the unabsorbed flux values.}
\label{fig:lightcurves}
\end{figure*}

In our WSRT observations we used the Multi Frequency Front Ends \citep{Tan1991} in combination with the IVC+DZB back end in continuum mode, with a bandwidth of 8x20\,MHz at all observing frequencies. Gain and phase calibrations were performed with the calibrator 3C 286 for all observations. The observations were analysed using the Multichannel Image Reconstruction Image Analysis and Display (MIRIAD; \citealt{Sault1995}) software package. There were multiple detections at 4.8\,GHz, while the 1.4\,GHz observation at 25\,days resulted in a non-detection. An observation at 1.4\,GHz with the Giant Metrewave Radio Telescope (GMRT), 11\,days after the burst, also resulted in a non-detection \citep{Chandra2014}. The radio data sets can be seen in Table \ref{tab:data} and the light curves are shown in Figure \ref{fig:lightcurves}. The 15.7\,GHz data exhibits signs of scintillation, noticeable at time scales of up to two weeks post GRB.

\subsection{Optical Afterglow Observations} \label{sec:obsoptical}
We observed the field of GRB\,140713A with the 2.5-m
Nordic Optical Telescope (NOT) equipped with ALFOSC
starting at 22:02 UT on 13 July 2014 \citep{Cano2014}. We obtained $5\times180$\,s frames in both r and i, and $5\times300$\,s in z. The NOT images have been calibrated to the USNO-B1 catalogue, using five stars in the field of view of GRB\,140713A. The B2-, R2- and I-band magnitudes of the five stars have been transformed into SDSS filters $r$, $i$, and $z$ (in the AB system) using the transformation equations in \citet{Jordi2006}. No object was detected within the XRT error circle of the GRB, and we find $3\sigma$ upper limits for an isolated point source in our images of $r > 24.30$, $i > 23.50$ and $z > 22.60$, at 0.1454, 0.1585 and 0.1738 days after the burst onset, respectively. The uncertainties associated with these upper limits are r = 0.16 mag, i = 0.15 mag, and z = 0.13 mag, which includes the standard deviation of the average offset between the instrumental and USNO-B1 catalogue magnitudes, and the variance of the transformation equations, which have been added in quadrature. The optical limits were converted into flux density using equation 3 in \citet{Frei1994}, followed by a transformation into Janskys. The flux density upper limits are shown in Figure \ref{fig:lightcurves}.

\subsection{Host galaxy observations and redshift determination} \label{sec:host}
While no optical afterglow detection was reported for this burst, the presence of a compact, non-varying source within the XRT circle with $R\sim24$\,mag was first noted by \citet{CastroTirado2014} and proposed as a potential host galaxy.

To test the likelihood of finding an unrelated galaxy within the XRT error circle of GRB\,140713A, we use the following relation \citep{Bloom2002}
\begin{align}
P_{\rm chance} = 1 - e^{-\pi r^{2}\sigma(\leq m_{R})}
\end{align}
where $r$ is the radius of the localization error circle and $\sigma(\leq m_{R})$ is the expected number of galaxies per arcsec$^{2}$ brighter than a given $R$-band magnitude limit. \citet{Bloom2002} state that if $P_{\rm chance} < 0.1$, the observed galaxy within the XRT error circle is most probably the host. Using the XRT error radius of 1.4 arcsec (90\% confidence) and $m_{R} = 24$\,mag we find $P_{\rm chance} = 0.028$, providing further evidence that this galaxy is the probable GRB host.

\begin{table}
\caption{Host galaxy photometry performed using a variety of filters and instruments. Note: magnitudes are not corrected for Galactic extinction; $E(B-V)=0.05$ \citep{Schlafly2011}.}
\centering
    \begin{tabular}{|r|c|l|} \\ \hline
    Filter & Magnitude (AB) & Instrument \\
    \hline
    $u$    &  24.30 $\pm$ 0.20 &  Keck/LRIS    \\
    $g$    &  24.22 $\pm$ 0.10 &  Keck/LRIS    \\
    $R$    &  24.00 $\pm$ 0.50 &  GTC/OSIRIS   \\
    $i$    &  23.11 $\pm$ 0.10 &  Keck/LRIS    \\
    $z$    &  22.49 $\pm$ 0.10 &  Keck/LRIS    \\
    3.6    &  21.45 $\pm$ 0.05 &  Spitzer/IRAC \\
    4.5    &  21.82 $\pm$ 0.05 &  Spitzer/IRAC \\ \hline
    \end{tabular}
\label{tab:host}
\end{table}

We obtained both imaging and spectroscopy of this source with the Low Resolution Imaging Spectrometer \citep[LRIS;][]{Oke1995} on the Keck~I 10m telescope at Maunakea, on the nights of 2014 August 30 and 31. Imaging was obtained in a variety of filters and totalled 480\,s in each of $U$-band, $g$-band and $i$-band, and 640\,s with the long-pass RG850 filter (similar to SDSS $z$-band). Images were reduced using the custom \texttt{LPIPE} pipeline and stacked.  Photometric calibration was performed using both Landolt standards acquired during the night and (for filters other than $U$) PS1 secondary standards in the field, and consistent results were obtained. Additionally, the source was observed by IRAC on-board the \textit{Spitzer} Space Telescope on 8 November 2016.  We performed aperture photometry on the source within IDL, using a 1.5 arcsec aperture for optical filters and a 2.4 arcsec aperture for IRAC. Photometry is provided in Table \ref{tab:host} and uncertainties are approximate, dominated by the photometric calibration.

\begin{figure}
\centering
\includegraphics[width=\linewidth]{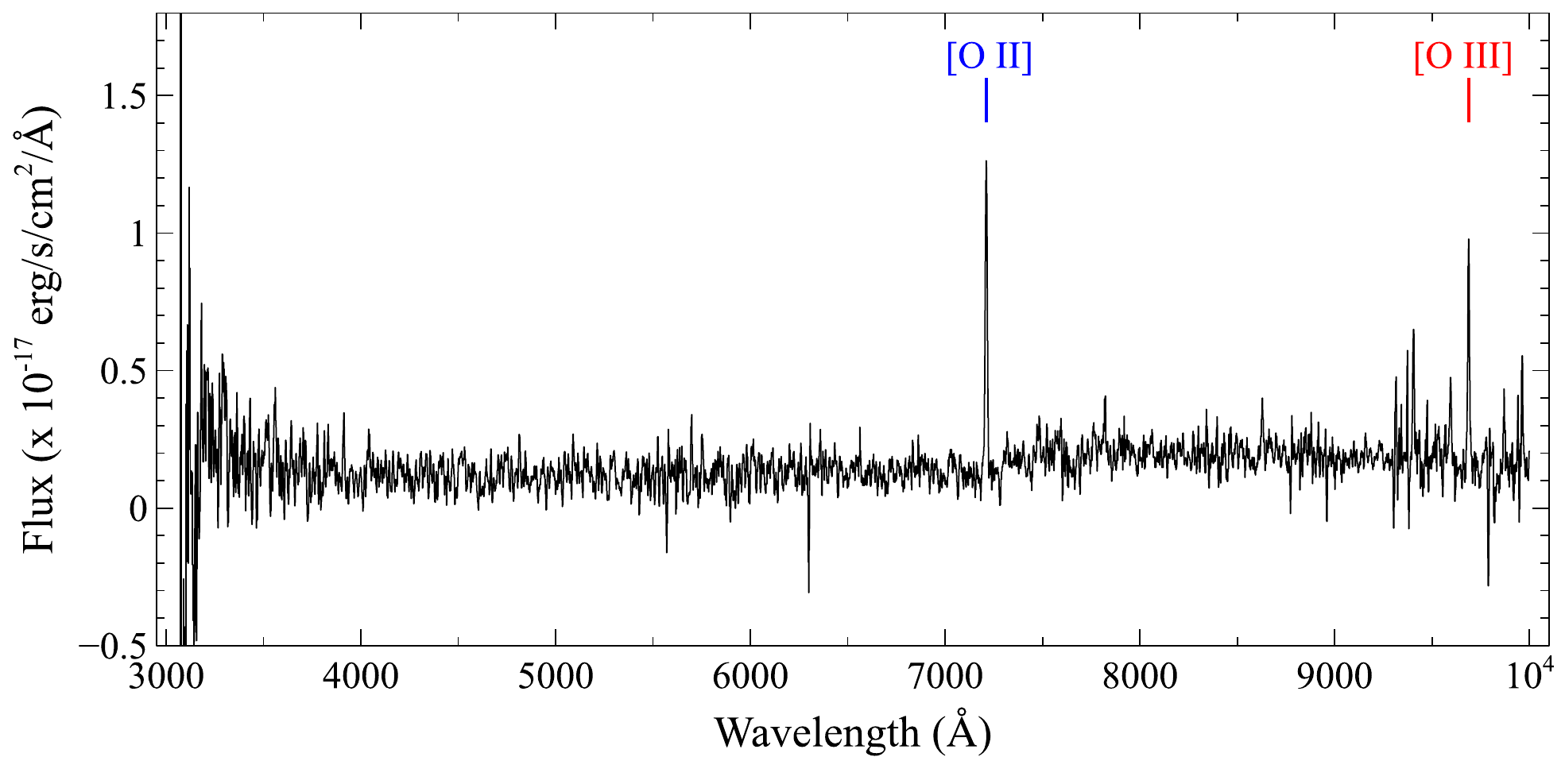}
\caption{The LRIS host galaxy spectrum. Highlighted are the 
[O II] doublet (blue) and the [O III] $\lambda$5007 (red) emission lines at a common redshift of $z=0.935$. The spectrum was smoothed with a 5 pixel boxcar for display purposes.}
\label{fig:hostspec}
\end{figure}

Our spectroscopic integration totalled approximately 1200 seconds (2$\times$600\,s blue, 2$\times$590\,s red) and employed the 400/3400 grism and 400/8500 grating on LRIS, covering a continuous wavelength range from the atmospheric cut-off to 10,290\,\AA. The reduced 1D spectrum (Figure \ref{fig:hostspec}) shows two strong emission features at wavelengths corresponding to [OII]$\lambda3727$ and [OIII]$\lambda5007$ at a common redshift of $z=0.935$, identifying this as the redshift of the system.

\begin{figure}
\centering
\includegraphics[width=\linewidth]{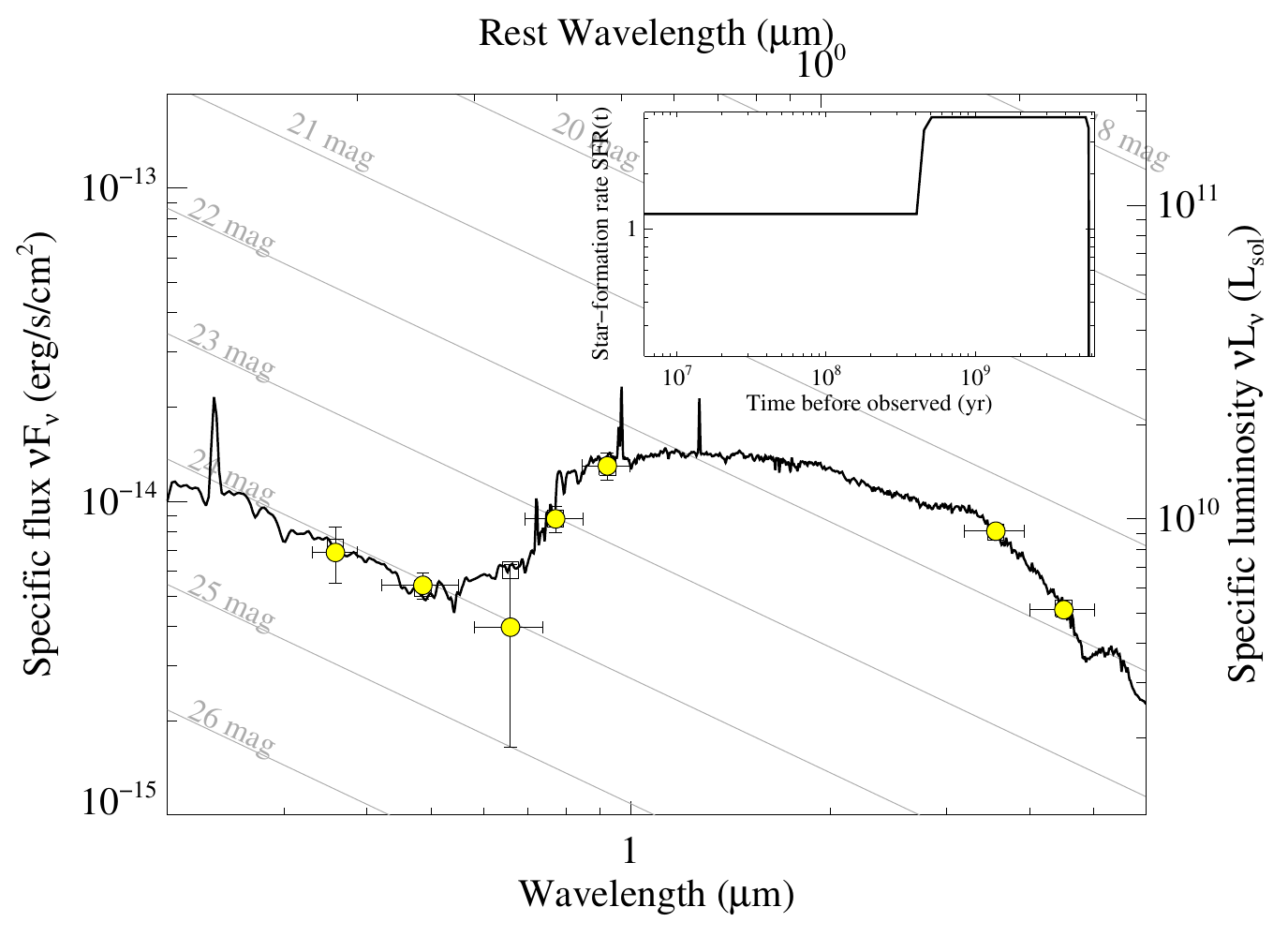}
\caption{SED of the potential host galaxy of GRB\,140713A. The data points are plotted in yellow (see Table \ref{tab:host}), the best-fitting model is denoted by the solid black line. The inset shows the assumed star-formation history prior to the galaxy redshift.}
\label{fig:hostsed}
\end{figure}

We performed an SED fit to our photometry (as well as the $R$ band point from \citealt{CastroTirado2014} - see Table \ref{tab:host}) using our custom SED analysis software \citep{Perley2013} assuming a star-formation history that is constant from $z=20$ to the observed redshift, except for an impulsive change at one point in the past. We found an excellent fit to a model with a total stellar mass of $2.2\times$10$^{10}$\,$M_\odot$ and a current star-formation rate of $1.2$\,$M_\odot$\,yr$^{-1}$ (see Figure \ref{fig:hostsed}). These values are typical of dark GRB hosts at similar redshift \citep{Perley2013} and generally of optically-selected galaxies at this epoch \citep{Contini2012}.

\subsection{X-ray Afterglow Observations} \label{sec:obsxray}
The \textit{Swift} satellite observed GRB\,140713A with the XRT starting at 18:45 UT on 13 July 2014, $\sim 80$ s after the \textit{Swift}/BAT trigger \citep{Mangano2014}. A coincident source was detected by the XRT and observations continued until $\sim 163$\,ks after the \textit{Swift}/BAT trigger for a total exposure time of 15.7\,ks.

\begin{figure}
\centering
\includegraphics[width=5.4cm,angle=270]{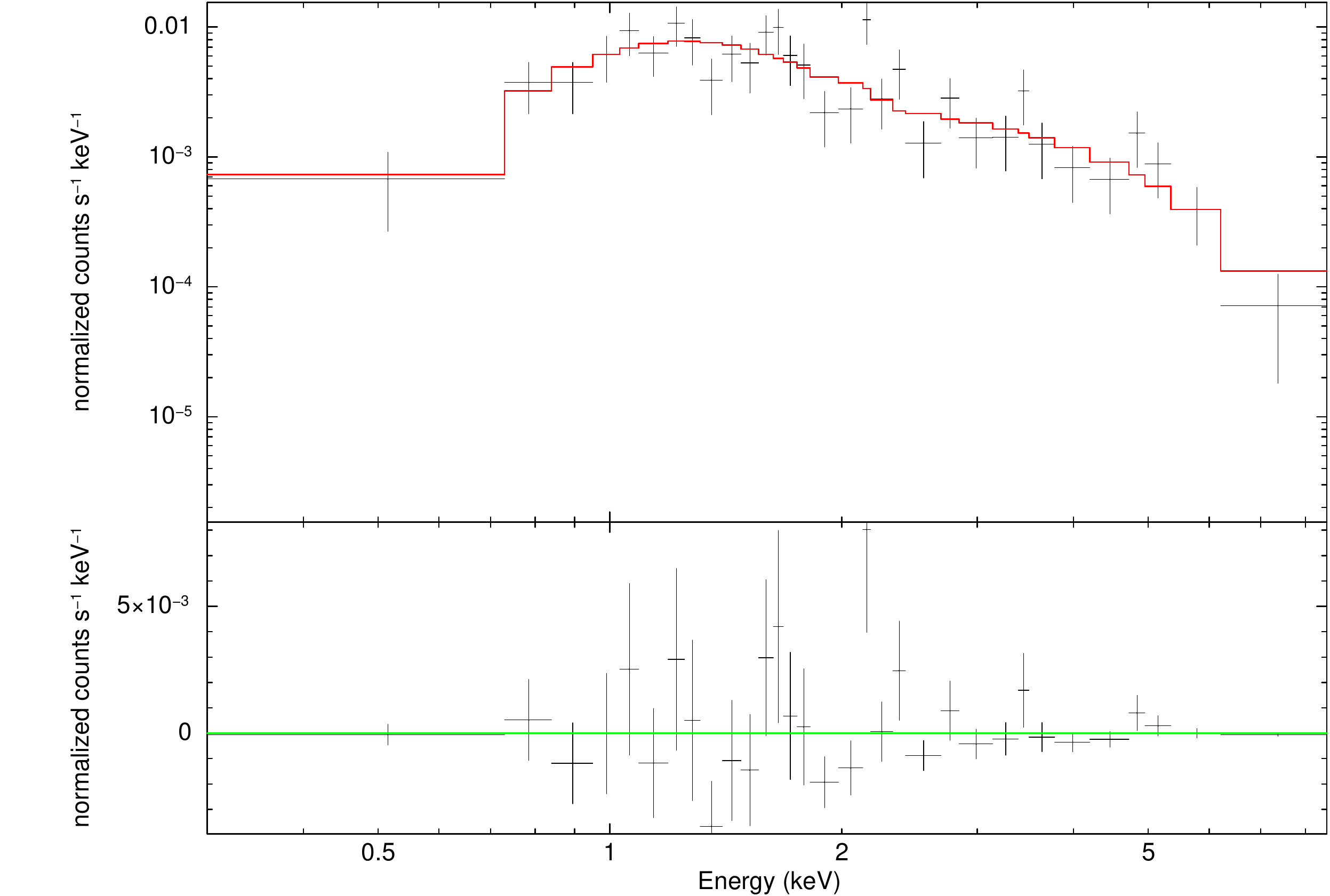}
\caption{Observed X-ray spectrum of GRB\,140713A from \textit{Swift}/XRT excluding the early time flaring data (see section \ref{sec:obsxray}). The absorbed power law model (red) was created using a Solar metallicity absorber.}
\label{fig:spectrum}
\end{figure}

The X-ray light curve of GRB\,140713A exhibits flaring $\sim 500$\,s after the \textit{Swift}/BAT trigger with a duration of $\sim 1000$\,s (taken from UKSSDC XRT GRB catalogue; \citealt{Evans2009}). The flaring most probably arises from extended central engine activity. As we only investigated emission from the afterglow we excluded the first 1500\,s of X-ray data in our modelling and only consider the later-time X-ray emission. We performed spectral analysis of the late-time spectral data using \textsc{xspec} (v12.9; \citealt{Arnaud1996}). We fit the data using an absorbed power law with a redshifted absorption component and a Galactic column density of $N_{\rm H, Gal}$ = $4.97\times10^{20}$\,cm$^{-2}$, calculated using the method described in \citet{Willingale2013}. Using a Solar metallicity absorber ($Z_{\odot}$) at $z=0.935$, we found a photon index $\Gamma=1.83^{+0.37}_{-0.33}$ and an excess intrinsic column density, $N_{\rm H, host}$ = $2.56^{+1.48}_{-1.12}\times10^{22}$\,cm$^{-2}$ (90\% confidence; C-stat = 114 for 155 degrees of freedom). GRB hosts typically have metallicities that differ from Solar metallicity \citep{Schady2012}, so we also fit the data using LMC-like ($Z_{\odot}/3$) and SMC-like ($Z_{\odot}/8$) metallicities at $z=0.935$ where $Z_{\odot}$ is solar metallicity. We found photon indices of $\Gamma=1.78^{+0.35}_{-0.32}$ and $\Gamma=1.74^{+0.34}_{-0.31}$, and intrinsic column densities of $N_{\rm H, host}$ = $5.10^{+3.07}_{-2.29}\times10^{22}$\,cm$^{-2}$ (90\% confidence; C-stat = 114 for 155 degrees of freedom) and $N_{\rm H, host}$ = $7.31^{+4.58}_{-3.36}\times10^{22}$\,cm$^{-2}$ (90\% confidence; C-stat = 114 for 155 degrees of freedom) for the LMC-like and SMC-like absorbers, respectively. The X-ray light curve and spectrum is shown in Figures \ref{fig:lightcurves} and \ref{fig:spectrum}. The XRT data and products were made available by the UK \textit{Swift} Science Data Centre (UKSSDC; \citealt{Evans2007,Evans2009}). The data were converted from unabsorbed flux into flux density at 2 keV using the photon index of $\Gamma = 1.83$ we obtained from the spectral analysis using a Solar metallicity absorber.

\section{Can we classify GRB140713A as a dark burst?} \label{sec:darkgrbs}
One criterion to determine if a GRB is indeed dark was proposed by \citet{Jakobsson2004}. They reported that an optical-to-X-ray spectral index $\beta_{\rm OX}<0.5$ at 11 hours would suggest that the GRB was optically sub-luminous with respect to the relativistic fireball model. The optical flux density is typically measured in the $R$ band and the X-ray flux density at 2 keV. This criterion was expanded to take into account the X-ray spectral information \citet{VanDerHorst2009}. Their criterion implies that a GRB can be classified as dark when $\beta_{\rm OX} - \beta_{\rm X} < - 0.5$ where $\beta_{\rm X}$ is the X-ray spectral index. We conducted a similar test, taking the unabsorbed X-ray flux at $\sim3.5$ hours (0.1454 days) and converted this into a flux density at 2 keV. We then calculated the spectral index between each optical band and the X-ray flux at 2\,keV. We find $\beta_{\rm OX}<0.5$ in all bands, well below the dark GRB threshold put forward by \citet{Jakobsson2004}. With $\beta_{\rm X} = 0.83^{+0.37}_{-0.33}$ (90\% confidence) we find that GRB\,140713A could also tentatively be classified as a dark GRB via the threshold described in \citet{VanDerHorst2009}. The results using both criterion are seen in Table \ref{tab:darkthreshold}.

\begin{table}
\caption{Spectral information using the optical upper limits and X-ray fluxes to determine if GRB\,140713A was dark using the thresholds described in section \ref{sec:darkgrbs}.}
\large
\centering
	\begin{tabular}{|l|c|c|}
    \hline
    Filter & $\beta_{\rm OX}$ & $\beta_{\rm OX} - \beta_{\rm X}$  \\ \hline
    $r$ & $<0.20$ & $<-0.30$ \\
    $i$ & $<0.26$ & $<-0.24$ \\
    $z$ & $<0.37$ & $<-0.13$ \\ \hline
    \end{tabular}
\label{tab:darkthreshold}
\end{table}

\section{Broadband afterglow modelling} \label{sec:modelling}
\subsection{Modelling method}
To investigate the origin of the optical darkness of GRB\,140713A we required an estimation of the optical flux we should expect to observe. Extrapolating the X-ray spectral index back to optical wavelengths implied that we should have observed an optical flux $\gtrsim 1$ order of magnitude brighter than the observed limits. To further investigate this discrepancy we modelled the afterglow data using the software package \textsc{boxfit} following the method described in \citet{vanEerten2012}. \textsc{boxfit} utilises the results of compressed radiative transfer hydrodynamic simulations to estimate the parameters of the expanding shock front and surrounding medium of a GRB using the downhill simplex method \citep{Nelder1965} optimised with simulated annealing \citep{Kirkpatrick1983}. Using \textsc{boxfit} as an alternative to the classical, analytical synchrotron models (i.e. \citealt{Granot2002}) allows us to fully compare the multi-wavelength data across a variety of times where the dynamical regimes of the afterglow change. \textsc{boxfit} models the afterglow from a single, initial injection of energy; we therefore omitted times for which flaring was observed in the X-ray data (see section \ref{sec:obsxray}). 

The afterglow model we used has nine parameters and is subsequently referred to as $\Phi$:
\begin{equation}
\Phi = [E_{\rm ISO},\,n,\,\theta_{\rm j/2},\,\theta_{\rm obs},\, p,\,\epsilon_{e},\,\epsilon_B,\,\xi_N,\,z]
\end{equation}
where $E_{\rm ISO}$ is the equivalent isotropic energy output of the blastwave, $n$ is the circumburst particle number density at a distance of $10^{17}$\,cm, $\theta_{\rm j/2}$ is the jet half-opening angle, $\theta_{\rm obs}$ is the observer angle with respect to the jet-axis, $p$ is the electron energy distribution index, $\epsilon_{e}$ and $\epsilon_B$ are the fractions of the internal energy in the electrons and shock-generated magnetic field, $\xi_N$ is the fraction of electrons that are accelerated, and $z$ represents the redshift. We assume a standard ${\rm \Lambda CDM}$ cosmology where $H_{\rm 0} = 68$\,km\,s$^{-1}$\,Mpc$^{-1}$ and $\Omega_{\rm M}$ = 0.31 \citep{Planck2016}, and calculate the corresponding luminosity distance, $d_{\rm L}$, from the redshift using the method described in \citet{Wright2006}.

Three of our nine model parameters - $z$, and therefore $d_{\rm L}$, $\theta_{\rm obs}$ and $\xi_{N}$ - are kept fixed. We justify these choices for the following reasons. The redshift, $z$, represents the distance to the GRB and is taken as the redshift of the host galaxy. For $\theta_{\rm obs}$ we assume that we observed the GRB on-axis, and we adopted $\xi_N=1$. This is to remove parameter degeneracies associated with these parameters - we do not have enough data to fully investigate the additional behaviour of these parameters. Our parameter ranges can be found in Table \ref{tab:modelpriors}. 

\begin{table}
\caption{The model parameter ranges for the afterglow model fitting.}
\large
\centering
\begin{threeparttable}
	\begin{tabular}{|l|c|c|c|}
    \hline
    Parameter & Min & Initial & Max \\ \hline
    $z^{\dagger}$ & - & 0.935 &  - \\
    $d_{\rm L}^{\dagger}$ (cm) & - & $1.92\times 10^{28}$ & - \\
    $E_{\rm ISO}$ (ergs) & $10^{47}$ & $10^{53}$ & $10^{56}$ \\
    $n$ cm$^{-3}$ & $10^{-5}$ & 1.0 & $10^{5}$ \\
    $\theta_{\rm j/2}$ (rad) & 0.01 & 0.1 & 0.5 \\
    $\theta_{\rm obs}^{\dagger}$ (rad) & - & 0 & - \\
    $p$ & 1.0 & 2.0 & 3.0 \\
    $\overline{\epsilon}_{e}$ & $10^{-5}$ & 0.1 & 1.0 \\
    $\epsilon_B$ & $10^{-10}$ & $10^{-5}$ & 1.0 \\
    $\xi_N^{\dagger}$ & - & 1.0 & - \\ \hline
    \end{tabular}
	\begin{tablenotes} \footnotesize
    \item{$^{\dagger}$ These parameters were frozen for the modelling. As we ran the models 	assuming the GRB was observed on axis ($\theta_{\rm obs}=0$) we set the azimuthal and radial resolution parameters to the recommended on-axis values of 1 and 1000, respectively.}
    \end{tablenotes}
\end{threeparttable}
\label{tab:modelpriors}
\end{table}

Figure \ref{fig:lightcurves} highlights that the late-time X-ray temporal slope was shallow; $-0.78(\pm0.09)$. For a rough estimate of $p$, we assume that the X-ray band is above the synchrotron cooling break frequency, which is most commonly observed at these times \citep{Curran2010,Ryan2015}, and use the following closure relations from \citet{Zhang2004}: the temporal relation ${\rm F_{\nu}}\propto t^{(2-3p)/4}$ and the spectral relation ${\rm F_{\nu}}\propto \nu^{-p/2}$ - to estimate $p$. We estimate from these relations that $1.6<p<1.8$ and $1.0<p<2.4$ for the temporal and spectral data respectively. These values suggest that the underlying electron distribution may be very low, i.e. $p < 2$. In light of this, we modified \textsc{boxfit} to allow fits where $p < 2$ by replacing $\epsilon_{e}$ with $\overline{\epsilon}_{e}$, where $\overline{\epsilon}_{e} = \epsilon_{e}(p-2)/(p-1)$ \citep{Granot2002}.

We ran \textsc{boxfit} for two different circumburst density environments - a homogeneous medium (subsequently labelled as ISM) and a stellar wind environment where the density decreases as $r^{-2}$, with $r$ the distance of the shock to the center of the stellar explosion. This allows us to test the significance of the environments on the best fitting models. The stellar wind environment was run under the medium-boosted wind setting of \textsc{boxfit}. We obtained the global best fit (i.e. lowest global $\chi^{2}$) for the data and calculated the partial derivatives around the best-fit values. We then used a bootstrap Monte Carlo (MC) method by perturbing the data set $10^{4}$ times within the flux errors to investigate the parameter distributions and confidence intervals (see \citealt{vanEerten2012} for a full discussion), showing the results in Figure \ref{fig:cornerplot}.

\subsection{Feasibility of parameter values and choice of environment for optical flux estimation}

\begin{figure*}
\centering
\includegraphics[width=\linewidth]{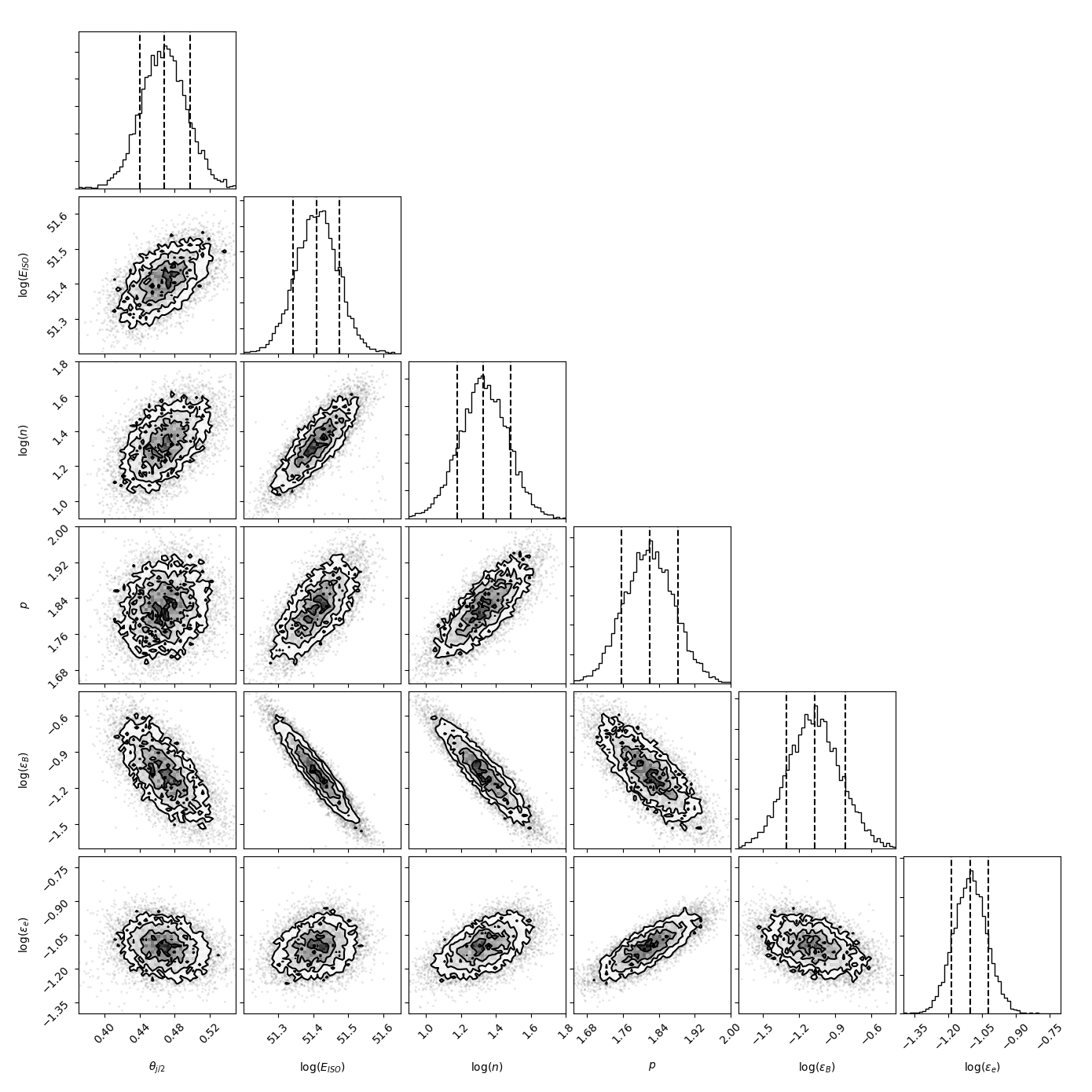}
\caption{Multi-dimensional projections of the parameter distributions derived using the MC bootstrap analysis around the \textsc{boxfit} ISM best fit values. The peak of the distributions and 68\% confidence intervals are shown.}
\label{fig:cornerplot}
\end{figure*}

\begin{table*}
\renewcommand{\arraystretch}{1.3}
\centering
\caption{The best fit parameter values derived from our MC analysis for two different circumburst density environments. The values quoted are the peak (median) of the MC distribution and 68\% ($1\sigma$) confidence intervals. The circumburst density is measured at a distance of $10^{17}$\,cm.}
	\begin{tabular}{|l|c|c|c|c|c|c|c|c|c|c|c|}
    \hline
    Environment & $E_{\rm ISO}$ (ergs) & $n$ (cm$^{-3})$ & $\theta_{\rm j/2}$ (rad) & $\theta_{\rm obs}^{\dagger}$ (rad) & $p$ & $\overline{\epsilon}_{e}$ & $\epsilon_B$ & $\xi_N^{\dagger}$ & $\chi^{2}_{r}$ \\ \hline
    ISM & $2.57^{+0.40}_{-0.33}\times10^{51}$ & $21.4^{+8.4}_{-5.8}$ & $0.47(\pm0.03)$ & 0 & $1.82(\pm0.06)$ & $7.88^{+1.42}_{-1.28}\times10^{-2}$ & $8.30^{+5.60}_{-3.30}\times10^{-2}$ & 1 & 4.21 \\
    Wind & $2.09^{+0.32}_{-0.25}\times10^{51}$ & $22.0^{+6.8}_{-3.1}$ & $0.51^{+0.04}_{-0.03}$ & 0 & $1.85^{+0.06}_{-0.05}$ & $5.64^{+1.37}_{-1.06}\times10^{-2}$ & $2.80^{+1.36}_{-1.01}\times10^{-1}$ & 1 & 4.70 \\ \hline
    \end{tabular}
\label{tab:boxfitresults}
\end{table*}

\begin{figure}
\centering
\includegraphics[width=\linewidth]{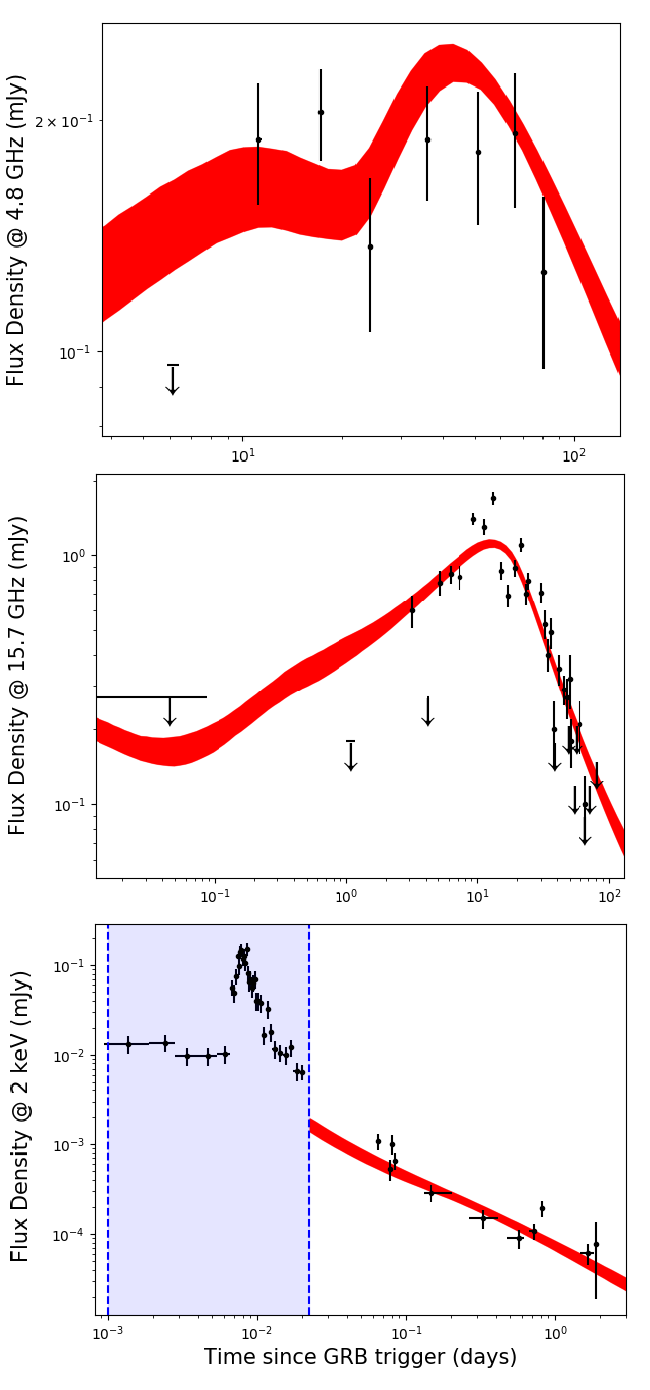}
\caption{Multi-wavelength light curves of GRB\,140713A for an ISM-like environment. The red region represents the 68\% confidence region. This was derived from generating model light curves from a random sample of 500 parameter sets, found using the MC bootstrap. We then plotted the model fluxes between the $16^{\rm th}$ and $84^{\rm th}$ percentiles (the distributions of fluxes in each model time bin were normal). The shaded blue region in the bottom window represents the X-ray flare that was omitted from the modelling.}
\label{fig:full_lc}
\end{figure}

Figure \ref{fig:cornerplot} shows that all the fitted parameters in our model, for an ISM-like environment, follow relatively normal and log-normal distributions. We see very similar results in the wind environment and both sets of parameter peak (median) and 68\% confidence ($1\sigma$) values are found in Table \ref{tab:boxfitresults}. There is some degeneracy between parameters, manifesting in the correlations we observe in Figure \ref{fig:cornerplot} (e.g. the positive correlation between $E_{\rm iso}$ and $n$ and anti-correlation between $\epsilon_{B}$ and $n$). Figure \ref{fig:full_lc} shows that our observations and models constrain the self-absorption and peak frequencies fairly well. Even with well-sampled optical light curves there can be correlations between parameters, because of the complexity and interdependence of several observable and physical parameters. However, since the characteristic synchrotron frequencies and peak flux are well constrained, our modelling work will provide a good estimated optical flux of GRB\,140713A. When comparing the best fit models of the two different circumburst density environments, the ISM fit has a lower global reduced $\chi^{2}$ statistic, $\chi^{2}_{r, \textsc{ism}} = 4.21$ compared to $\chi^{2}_{r, \textsc{wind}} = 4.70$, but this difference is not statistically significant. Both models fail to reproduce several early time non-detections in the 4.8 and 15.7\,GHz light curve, but at these times scintillation is clearly visible in the data and produces significant short-term flux variability. Both environments produce consistent values within $1\sigma$ for $E_{\rm ISO}$, $n$, $\theta_{j/2}$, $\overline{\epsilon}_{e}$ and $p$; and consistent within $2\sigma$ for $\epsilon_{B}$.

In both the ISM and wind case, our \textsc{boxfit} models prefer a large jet half-opening angle ($\theta_{\rm j/2}\sim0.5$\,rad) implying that a jet break would occur at $\sim 25-30$\,days post-GRB (calculated using equation 3 in \citealt{Starling2009}, see also \citealt{Frail2001}). This is clearly visible in the 4.8\,GHz band light curve in Figure \ref{fig:full_lc}. GRBs exhibit a range of jet opening angles; ranging from the narrow ($\theta_{\rm j/2} \lesssim 0.1$\,rad; \citealt{Frail2001,Ryan2015}) to the wide (e.g., GRB\,970508; \citealt{Frail2000} or GRB\,000418; \citealt{Panaitescu2002}), with our jet half-opening angle estimation comfortably sitting within the distribution. Both circumburst environments also preferred a scenario with a hard electron energy distribution with $p\sim 1.85$. Although low $p$ values for these two environments have been derived for other GRBs as well, they are in the tails of GRB parameter distributions (e.g. \citealt{Curran2010,Ryan2015}).

With the hard electron energy distribution ($p < 2$) preferred by the afterglow modelling, we exercised caution when interpreting the physical meaning of $\overline{\epsilon}_{e}$ - introduced to allow \textsc{boxfit} to model fits where $p < 2$. For $p > 2$ where one can fit for $\epsilon_{e}$, you can simply estimate the energy of the shocked electrons from the following relation
\begin{align}
E_{e} = \epsilon_{e} E_{int} \label{eq:pgtr2}
\end{align}
where $E_{e}$ is the energy density of the shocked electrons and $E_{int}$ is the energy density of the post-shock fluid. However, as we had a best fit where $p < 2$ we had to account for the upper energy cut-off of the electron energy distribution by using the following relation
\begin{align}
E_{e} = \overline{\epsilon}_{e} E_{int} \frac{(p-1)}{(p-2)} \left[1-\left(\frac{\gamma_M}{\gamma_m}\right)^{2-p}\right] \label{eq:plt2}
\end{align}
where $\gamma_{M}$ and $\gamma_{m}$ represent the maximum (cut-off) and minimum Lorentz factors accounting for the cut-off with $\overline{\epsilon}_{e}$ defined as before \citep{Granot2002}. For further details on $\overline{\epsilon}_{e}$, see Appendix \ref{ap:energydensity}. If we assume values of $\gamma_{M} \sim 10^{7}$ and $\gamma_{m} \sim 10^{3}$, and take the derived values for $p$ and $\overline{\epsilon}_{e}$ for both the ISM and wind environments from Table \ref{tab:boxfitresults}, we find that the typical energy densities from equation \ref{eq:pgtr2} are a factor of $\sim 15 - 20$ lower than for equation \ref{eq:plt2}. This is not surprising given the form of equation \ref{eq:plt2} - as $p \rightarrow 2$ the energy of the electrons using the above relation asymptotically scales towards infinity, resulting in energy efficiencies, $\epsilon_{e}$ > 1, which are not physical. In section \ref{sec:modelling} we discussed our model parameter selection, including fixing $\xi_{\rm N} = 1$ as we do not have sufficient data to explore the degeneracy of this parameter with the other parameters. A linear decrease in $\xi_{\rm N}$ would result in a linear increase in energy, $E_{\rm ISO}$ but simultaneous linear decreases in both $\epsilon_{e}$ and $\epsilon_{B}$ \citep{Eichler2005}. Therefore, if we had set a value of $\xi_{\rm N} = 0.1$, we would have seen an increase in the available energy budget to $E_{\rm ISO}\sim 10^{52}$\,ergs but also would have seen $\epsilon_{B} \sim 0.01$ and $\epsilon_{e} \sim 0.1$ for both the ISM and wind environments, in which case $\epsilon_{e}$ would be physical.

\section{Optical Darkness} \label{sec:opdark}
The temporal and spectral behaviour of GRB\,140713A in the gamma-ray, X-ray and radio regimes are very typical of GRB afterglows. Our optical observations were taken at such early times and with a sensitivity that a counterpart should have been detected. In section \ref{sec:host} we discussed observations of a associated host galaxy in a number of optical and near-infrared bands. The 1D spectrum shows two strong emission lines - [OII]$\lambda3727$ and [OIII]$\lambda5007$ - both occurring at the same redshift, $z = 0.935$. We therefore rule out that the optical darkness is due to low intrinsic luminosity of the GRB or a high-redshift nature.

\begin{figure}
\centering
\includegraphics[width=\linewidth]{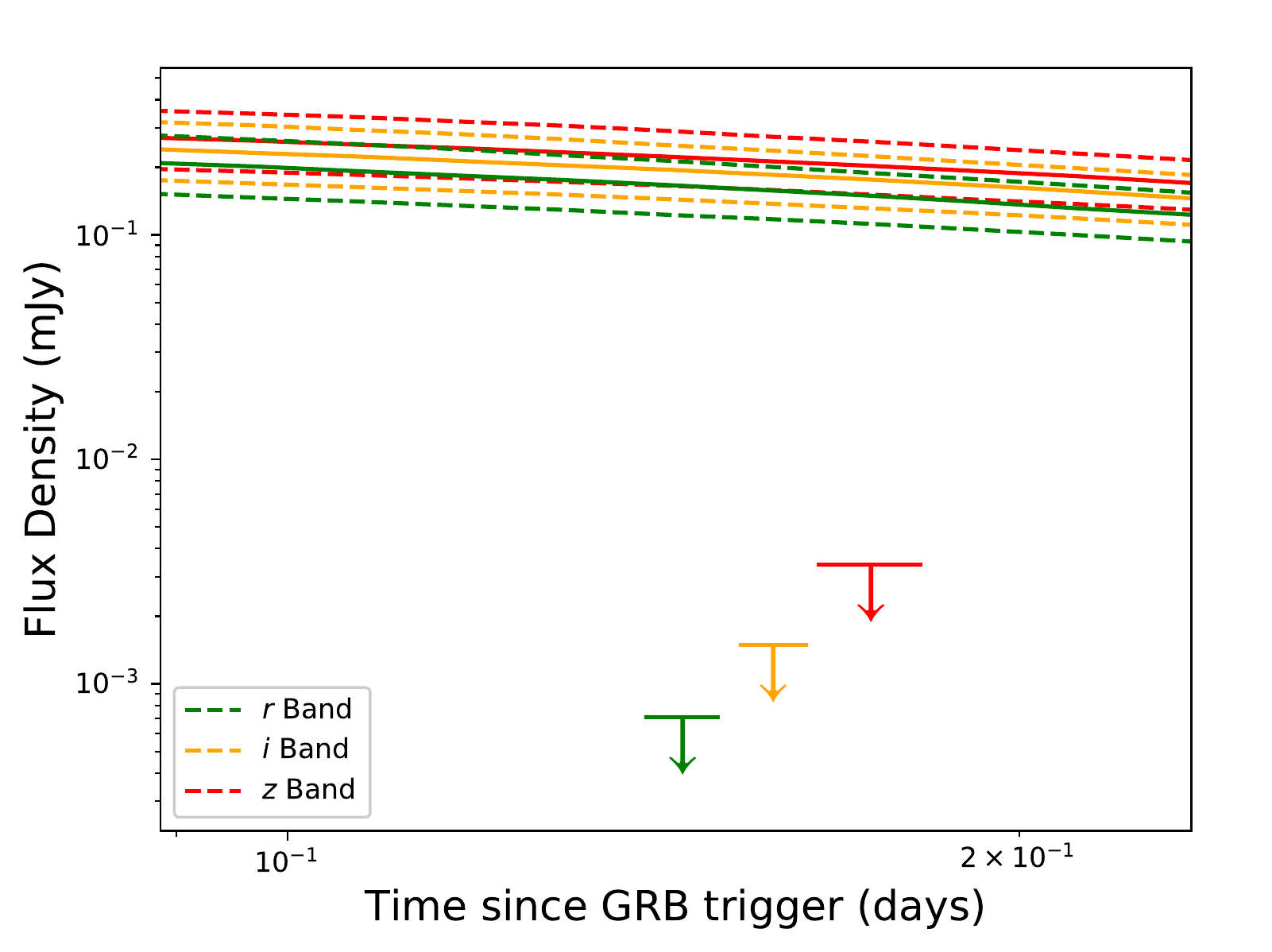}
\caption{Optical light curves of the $r$ (green), $i$ (orange) and $z$ (red) bands for an ISM-like model. The solid lines represent the $50^{\rm th}$ percentile (median) values and the dashed lines represent the $16^{\rm th}$ and $84^{\rm th}$ percentile values (the 68\% confidence intervals).}
\label{fig:opticallc}
\end{figure}

As the physical parameters of both environments were very similar, but the ISM environment had a smaller $\chi^{2}_{r}$ derived from our \textsc{boxfit} modelling, we used the ISM derived parameters to estimate the optical flux in the $r$, $i$ and $z$ bands. We randomly sampled 500 parameter sets from the $10^{4}$ sets derived from the MC bootstrap and produced light curves in the $r$, $i$ and $z$ bands. The flux values for each of the 500 light curve models, in each time bin, were found to follow log-normal distributions (see appendix \ref{ap:opticaldists} for examples). We therefore plotted the $16^{\rm th}$, $50^{\rm th}$ and $84^{\rm th}$ percentiles of each time bin to illustrate the most-probable flux and 68\% confidence intervals from our model (see Figure \ref{fig:opticallc}). We also quote these percentiles as our results in Table \ref{tab:extinction}. These were then compared to the expected fluxes to our observational upper limits. The estimated flux in the $r$, $i$ and $z$ bands are $\gtrsim 2$ orders of magnitude above the upper limits (Table \ref{tab:extinction}). The estimated flux values from Figure \ref{fig:opticallc} confirm that optical observations were promptly taken and should have led to detections of the GRB\,140713A counterpart. The remaining plausible explanation for the optical darkness of this GRB is optical extinction in the line of sight towards the source.

\begin{table*}
\renewcommand{\arraystretch}{1.2}
\centering
\caption{Table containing the observed optical limits, \textsc{boxfit} fluxes, required level of extinction and derived host extinctions ($A^{\rm host}_{V}$) using Milky Way-like, LMC-like and SMC-like extinction models. The lower limits shown for the required level of extinction were calculated from the magnitude difference between the observational upper limit and the faintest estimate from \textsc{boxfit} at the given wavelength (i.e. 18.6\,mag for $r$ band). Host extinction values have been corrected for Galactic extinction - $E(B-V)=0.05$\,mag; $A^{\rm host}_{V} = 0.16$\,mag; \citep{Schlafly2011}. Quoted errors on \textsc{boxfit} fluxes at 68\% confidence.}
	\begin{tabular}{|r|c|c|c|c|c|c|c|c|c|}
    \hline
    Filter & $\Delta T$ & Mag & Flux & \textsc{boxfit} Flux & \textsc{boxfit} Mag & Req Ext & Galactic & LMC & SMC \\ 
     & (Days) & (AB) & (mJy) & (mJy) & (mag) & (mag) & $A^{\rm host}_{V}$ & $A^{\rm host}_{V}$ & $A^{\rm host}_{V}$ \\
     & & & & & & & (mag) & (mag) & (mag) \\ \hline
    $r$ & 0.15 & > 24.3 & $<7.1\times10^{-4}$ & $0.17^{+0.05}_{-0.04}$ & $18.3(\pm0.3)$ & $>5.7$ & $>3.1$ & $>3.2$ & $>2.9$ \\
    $i$ & 0.16 & > 23.5 & $<1.5\times10^{-3}$ & $0.19(\pm0.05)$ & $18.2(\pm0.3)$ & $>5.0$ & $>3.2$ & $>3.3$ & $>3.0$ \\
    $z$ & 0.17 & > 22.6 & $<3.4\times10^{-3}$ & $0.20^{+0.06}_{-0.05}$ & $18.1^{+0.4}_{-0.2}$ & $>4.1$ & $>3.2$ & $>3.2$ & $>3.1$ \\ \hline
    \end{tabular}
\label{tab:extinction}
\end{table*}

The estimated optical flux values are given in Table \ref{tab:extinction} and allowed us to uncover the potential source of optical extinction. We used the \textsc{boxfit} derived fluxes to estimate lower limits on the extinction in the $r$, $i$ and $z$ bands, ranging from 4.2 to 5.7\,mag. We used Milky Way-like ($R_{V} = 3.1$; \citealt{Cardelli1989}), LMC-like ($R_{V} = 3.41$; \citealt{Gordon2003}) and SMC-like ($R_{V} = 2.74$; \citealt{Gordon2003}) extinction models to derive the required host extinction ($A^{\rm host}_{V}$) after transforming the observed bands into their corresponding wavelengths in rest frame of the host galaxy at $z = 0.935$. We then subtracted the Galactic extinction contribution $E(B-V)^{\rm Gal} = 0.05$\,mag. The three extinction models produced similar results; see Table \ref{tab:extinction}. The most constraining limit from the Milky Way-like extinction model was $A^{\rm host}_{V} > 3.2$\,mag equivalent to $E(B-V)^{\rm host} > 1.0$\,mag.

We independently estimated the host extinction level using the relationship between X-ray absorption and optical extinction \citep{Gorenstein1975,Predehl1995}. A more recent study published in 2009 constrained this relationship to $N{\rm_H}$ (cm$^{-2}$) = $2.21(\pm0.09)\times10^{21} A_{V}$ \citep{Guver2009}. In section \ref{sec:obsxray} we estimated the intrinsic hydrogen column density of GRB\,140713A, $N_{\rm H, host}$ = $2.6^{+1.48}_{-1.12}\times10^{22}$\,cm$^{-2}$ (90\% confidence, assuming Solar metallicity absorber). We estimate the expected optical host extinction based on the relation between extinction and X-ray column density for the Milky Way, LMC and SMC in \citet{Guver2009}. We calculate that the extinction for the host is $A^{\rm host}_{\rm V} = 11.6^{+7.5}_{-5.3}$\,mag (90\% confidence), equivalent to $E(B-V)^{\rm host} = 3.7^{+2.4}_{-1.7}$\,mag for the Milky Way-like extinction model. Combining the SMC-like and LMC-like absorber intrinsic column densities derived in section \ref{sec:obsxray} with the relation from \citet{Guver2009}, results in estimated host extinction values of $A_{V} = 23.1^{+13.9}_{-10.4}$\,mag and $A_{V} = 33.1^{+20.7}_{-15.2}$\,mag, respectively. Our estimated host extinction using the hydrogen column density is in good agreement with the extinction limits calculated from the \textsc{boxfit} generated light curves, and suggests that the source of the optical extinction is due to dust within the host galaxy.

\subsection{Comparing the extinction of dark GRBs}
Only a handful of dark GRBs with accompanying radio data have been observed. The explosion and circumburst properties of these GRBs were compared in \citet{Zauderer2013}. Table \ref{tab:darkgrbs} summarises the estimated host extinction of all of these bursts to date and GRB\,140713A from this investigation.

\begin{table}
\footnotesize\setlength{\tabcolsep}{2.0pt}
\renewcommand{\arraystretch}{1.2}
\caption{Required host extinction values for a number of dark GRBs with complementary radio data. All extinctions are quoted directly from their respective sources unless otherwise stated and are displayed in the rest frame of the host.}
\centering
	\begin{threeparttable}
	\begin{tabular}{|l|c|c|}
    \hline
    GRB Name& $A^{\rm host}_{V}$ & Reference \\
     & (mag) & \\ \hline
    970828 & $>3.8$ & \citealt{Djorgovski2001} \\
    000210 & $0.9-3.2$ & \citealt{Piro2002} \\
    020809 & $0.6-1.5$ & \citealt{Jakobsson2005} \\
    051022 & $>8.2^{a}$ & \citealt{Rol2007} \\
    110709B & $>5.3$ & \citealt{Zauderer2013} \\
    111215A & $>7.5$ & \citealt{VanDerHorst2015} \\
    140713A & $>3.2$ & This work \\ \hline
    \end{tabular}
    \begin{tablenotes}
    \item{$^{a}$: Most constraining limit derived using their quoted $J$ band extinction in the host rest frame ($A^{\rm host}_{J}>2.3$\,mag) and transforming this into the $V$ band extinction assuming a Milky Way-like extinction curve.}
    \end{tablenotes}
    \end{threeparttable}
    \label{tab:darkgrbs}
\end{table}

The required host extinction values vary significantly in this small sample from modest ($A^{\rm host}_{V} \lesssim 1.5$\,mag) to high ($A^{\rm host}_{V} > 8.2$\,mag) and GRB\,140713A is typical among the other dark GRBs. Interestingly, at least five of the seven GRBs exhibit required extinctions of $>3$\,mag. The levels of extinction are in good agreement with larger sample studies of optically dark GRB host galaxies \citep{Perley2009,Perley2013}. The results therefore suggest that the optical extinction of a significant fraction of dark GRBs is at least partially due to dust-obscuration in the host galaxy, either in the local environment of the progenitor or throughout the galaxy.   

\section{Conclusions} \label{sec:conclusions}
The afterglow of GRB\,140713A was detected in both the X-ray and radio bands but not seen to deep limits in optical and near-infrared observations. We measured the likely host galaxy redshift of $z=0.935$, allowing us to rule out a high-redshift origin. 
We investigated the origin of optical darkness in this GRB utilising hydrodynamical jet simulations through the modelling software \textsc{boxfit}. We produced a number of models in both an ISM-like and wind circumburst environment to estimate what level of optical flux we could have expected from the afterglow. The models provided good fits to the observed data preferring a wide jet half-opening angle ($\theta_{\rm j/2} \sim 0.5$\,rad) and a hard electron energy distribution ($p \sim 1.85$). Crucially, the models predicted that the observed optical afterglow should have been $\sim 2$ orders of magnitude brighter than our observed upper limits and therefore easily observable, ruling out an intrinsically low luminosity optical afterglow. From the discrepancy between the estimated optical flux values and our observations we estimated that we require an extinction $A_{V}^{\rm host} > 3.2$\,mag in the rest frame of the host. The host optical extinction, inferred from the hydrogen column density measured in the X-ray afterglow spectra data, was consistent with our requirements. 
We therefore conclude that the optical darkness of GRB\,140713A is most likely caused by a large amount of extinction either in the local vicinity of the progenitor or throughout the host galaxy.

\section{Acknowledgements}
We greatly appreciate the support from the WSRT and AMI staff in their help with scheduling and obtaining these observations. The WSRT is operated by ASTRON (Netherlands Institute for Radio Astronomy) with support from the Netherlands foundation for Scientific Research. The AMI arrays are supported by the University of Cambridge and the STFC. Some of the data presented herein were obtained at the W.M. Keck Observatory, which is operated as a scientific partnership among the California Institute of Technology, the University of California and the National Aeronautics and Space Administration (NASA); the Observatory was made possible by the generous financial support of the W.M. Keck Foundation. The Nordic Optical Telescope is operated on the island of La Palma by the Nordic Optical Telescope Scientific Association in the Spanish Observatorio del Roque de los Muchachos of the Instituto de Astrofisica de Canarias.
This work made use of data supplied by the UK Swift Science Data Centre at the University of Leicester.
ABH is supported by a STFC studentship. RLCS, KW and NRT acknowledge support from STFC.
GEA acknowledges the support of the European Research Council Advanced Grant 267697 `4 Pi Sky: Extreme Astrophysics with Revolutionary Radio Telescopes'. GEA is the recipient of an Australian Research Council Discovery Early Career Researcher Award (project number DE180100346) funded by the Australian Government.
We thank the referee for their constructive feedback which improved the paper.

\bibliographystyle{mnras}
\bibliography{grb140713a}

\appendix

\section{Interpretation of a hard electron energy distribution} \label{ap:energydensity}
We assume a power-law accelerated electron number density according to $n_{e}(\gamma_{e}) = C \gamma_{e}^{-p}$ between lower cut-off $\gamma_{m}$ and upper cut-off $\gamma_{M}$, where $\gamma_{e}$ the Lorentz factor of individual electrons in the frame locally co-moving with the fluid, and $C$ a constant of proportionality constrained by the total number density of electrons. Following \citet{Granot2002}, we use $\overline{\epsilon}_{e}$ rather than $\epsilon_{e}$ as a fit parameter to model the fraction of available blast wave energy that resides in the accelerated electron population:
\begin{align}
\gamma_{m} \equiv \frac{\overline{\epsilon}_{e} E_{int}}{\xi_{N} n m_{e} c^{2}} \equiv \frac{(p-2)}{(p-1)} \frac{\epsilon_{e} E_{int}}{\xi_{N} n m_{e} c^{2}}.
\end{align}
Here $E_{int}$ is the internal post-shock energy density of the fluid and $n$ its post-shock number density. The upper cut-off $\gamma_{M}$ reflects the balance between shock-acceleration time and synchrotron loss time. We do not account for $\gamma_{M}$ when generating light curves, given that its observational signature (an exponential drop in flux) will lie orders of magnitude above the X-ray band for reasonable model parameter values. If $p > 2$, $\gamma_{M}$ can also be ignored when inferring the total energy available to electrons $E_{e}$ from our fit result for $\overline{\epsilon}_{e}$, according to $E_{e} = \epsilon_{e} E_{int} = \overline{\epsilon}_{e} E_{int} (p-1)/(p-2)$.

More generally, when allowing for $p < 2$ as well, we have
\begin{align}
\begin{split}
E_{e} = C m_{e} c^{2} \int_{\gamma_{m}}^{\gamma_{M}} d \gamma_{e} \gamma_{e}^{1-p} \\
&\hspace*{-94pt} \approx \left\{\begin{array}{ll} \gamma_{m}^{2} n_{e} (\gamma_{m}) m_{e} c^{2} / (p-2), & p > 2, \\ \gamma_{m}^{2} n_{e} (\gamma_{m}) m_{e} c^{2} \ln \left[ \gamma_{M} / \gamma_{m} \right], & p = 2 \\ \gamma_{M}^{2} n_{e}(\gamma_{M}) m_{e} c^{2} / (2-p), & p < 2. \\  \end{array}\right.
\end{split}
\end{align}
Here the $p > 2$ and $p < 2$ cases have their energy estimate dictated by $\gamma_{m}$ and $\gamma_{M}$ respectively (with $\gamma_{M}$ and $\gamma_{m}$ respectively being ignored in the preceding equations). If all terms are accounted for, and $E_{e,old}$ is our inferred electron energy when ignoring $\gamma_{M}$, the actual value for $E_{e}$ is given by the following relation
\begin{align}
\begin{split}
E_{e} = E_{e,old} \left[1 - \left(\frac{\gamma_{M}}{\gamma_{m}} \right)^{p-2} \right] \\ &\hspace*{-96pt} = \overline{\epsilon}_{e} \frac{(p-1)}{(p-2)} E_{int} \left[ 1 - \left( \frac{\gamma_{M}}{\gamma_{m}} \right)^{p-2} \right].
\end{split}
\end{align}

\section{Optical flux distributions derived from the MC samples} \label{ap:opticaldists}
\begin{figure}
\centering
\includegraphics[width=\linewidth]{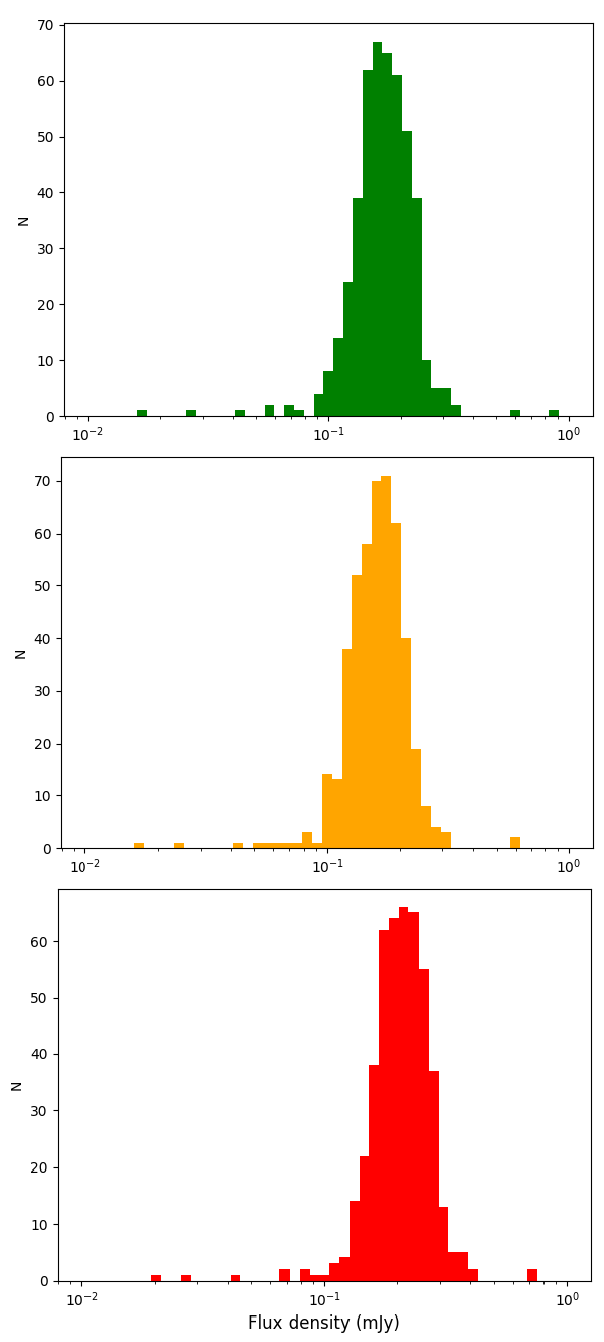}
\caption{Optical flux distribution of the $r$ (green), $i$ (orange) and $z$ (red) bands. The time bins represented in each window are 0.15\,days for $r$ band, 0.16\,days for $i$ band, and 0.17\,days for $z$ band.}
\label{fig:opticaldist}
\end{figure}

Figure \ref{fig:opticaldist} represents the flux distributions from the 500 randomly sampled parameter set light curves discussed in section \ref{sec:modelling}. The time bin of the displayed fluxes in the $r$, $i$, and $z$ bands correspond to the times of the observations in those filters (see Table \ref{tab:extinction}). The distributions are clearly log-normal so the median flux values and 68\% confidence intervals can be quoted using the $16^{\rm th}$, $50^{\rm th}$ and $84^{\rm th}$ percentiles. 

\bsp	
\label{lastpage}
\end{document}